\documentclass[10pt]{iopart}

\usepackage{graphicx,latexsym,color}
\usepackage{cite,citesort}

\usepackage{amsgen,amsfonts,amsbsy,amssymb}

\usepackage{listings}

\usepackage{bm}
\usepackage{tikz}
\usepackage{pgfplots}

\newcommand{\R}{\mathbb{R}}
\newcommand{\C}{\mathbb{C}}

\newcommand{\ie}{\textit{i.e.}\/, }
\newcommand{\eg}{\textit{e.g.}\/, }
\newcommand{\cf}{\textit{cf.}\/, }

\providecommand*{\mrm}[1]{\mathrm{#1}}
\providecommand*{\unit}[1]{\ensuremath{\mrm{\,#1}}}
\providecommand*{\eu}{\ensuremath{\mrm{e}}}
\providecommand*{\iu}{\ensuremath{\mrm{i}}}

\providecommand*{\micro}{\ensuremath{\mrm{\mu}}}
\providecommand*{\degree}{\ensuremath{^\circ}}

\newcommand{\minimize}{\mrm{minimize}}

\newcommand{\subto}{\mrm{subject\ to}}

\def\Xint#1{\mathchoice
   {\XXint\displaystyle\textstyle{#1}}%
   {\XXint\textstyle\scriptstyle{#1}}%
   {\XXint\scriptstyle\scriptscriptstyle{#1}}%
   {\XXint\scriptscriptstyle\scriptscriptstyle{#1}}%
   \!\int}
\def\XXint#1#2#3{{\setbox0=\hbox{$#1{#2#3}{\int}$}
     \vcenter{\hbox{$#2#3$}}\kern-.5\wd0}}

\def\dashint{\Xint-}

\begin{document}

\title{On the physical limitations for radio frequency absorption in gold nanoparticle suspensions}

\author{Sven~Nordebo$^1$, Mariana~Dalarsson$^1$, Yevhen~Ivanenko$^1$, Daniel~Sj\"{o}berg$^2$, Richard~Bayford$^3$}

\address{$^1$ Department of Physics and Electrical Engineering, Linn\ae us University, 351 95 V\"{a}xj\"{o}, Sweden.
E-mail: \{sven.nordebo,mariana.dalarsson,yevhen.ivanenko\}@lnu.se.}
\address{$^2$ Department of Electrical and Information Technology, Lund University, Box 118, 
221 00 Lund, Sweden. E-mail: daniel.sjoberg@eit.lth.se.}
\address{$^3$ The Biophysics and Bioengineering Research Group, Middlesex University, Hendon campus, 
The Burroughs, London, NW4 4BT, United Kingdom. E-mail: R.Bayford@mdx.ac.uk.}

\maketitle
\ioptwocol

\begin{abstract}
This paper presents a study on the physical limitations for radio frequency absorption in gold nanoparticle suspensions.
A canonical spherical geometry is considered consisting of a spherical suspension of colloidal gold nanoparticles 
characterized as an arbitrary passive dielectric material which is immersed in an arbitrary lossy medium.
A relative heating coefficient and a corresponding optimal near field excitation are defined taking the skin effect of the 
surrounding medium into account.  For small particle suspensions the optimal excitation
is an electric dipole field for which explicit  asymptotic expressions are readily obtained. It is then proven that the optimal
permittivity function yielding a maximal absorption inside the spherical suspension is a conjugate match with respect to the surrounding lossy material.
For a surrounding medium consisting of a weak electrolyte solution the optimal conjugate match can then readily be realized at a 
single frequency, \eg by tuning the parameters of a Drude model corresponding to the electrophoretic particle acceleration mechanism. 
As such, the conjugate match can also be regarded to yield an optimal plasmonic resonance.
Finally, a convex optimization approach is used to investigate the realizability of a passive material to approximate the desired conjugate match over a finite bandwidth.
The relation of the proposed approach to general Mie theory as well as to the approximation of metamaterials are discussed.
Numerical examples are included to illustrate the ultimate potential of heating in a realistic scenario in the microwave regime.
\end{abstract}

\section{Introduction}
A number of publications have proposed that biological tissue can be heated quickly and selectively by the use of gold nanoparticles (GNPs)
that are subjected to a strong time-harmonic electromagnetic field, \eg at 13.56\unit{MHz} or in the microwave regime, where the band is chosen due to regulations,
see \eg \cite{Collins+etal2014,Sassaroli+etal2012,Cardinal+etal2008,Gannon+etal2008,Moran+etal2009,marquez2013hyperthermia}.
The aim is to develop a method with the potential to treat cancer.
The GNPs can also be used as a contrast agent for electrical impedance tomography, particularly when combined with tumour targeting \cite{Callaghan+etal2010}.
The fundamental idea relies on the fact that ligands can be attached to the GNPs to target cancer cells.
The rapid rate of growth of the cancer cells causes them to intake an abnormal amount of nutrients and
the GNPs can hence be coated with folic acid to target the bio-markers or antigens that are highly specific to the cancer cells, see \eg \cite{Dreaden+etal2012}.
The aim of the radio frequency (RF) treatment is then to promote local cell death only in the cancer,
in addition to increasing the pore size to improve delivery of large-molecule chemotherapeutic and immunotherapeutic agents.
The sensitivity of the cancer cells to elevated temperatures enables the tumor growth to be slowed or stopped by transient heating to 40-46\unit{\degree C}
for periods of 30 minutes or more, whilst also increasing the tumor sensitivity to chemotherapy and radiotherapy \cite{Hildebrandt+etal2002}.

The physical explanation to the RF-heating of the GNPs is not fully understood and there are many phenomenological hypotheses
proposed to explain the heating \cite{Collins+etal2014,Sassaroli+etal2012,Corr+etal2012,Cardinal+etal2008,Gannon+etal2008,Moran+etal2009}.
Recently, it has also been questioned whether metal nanoparticles can be heated in radio frequency at all, and
several authors have been unable to find neither theoretical nor experimental results to support the hypothesis,
see \eg \cite{Collins+etal2014,Gupta+etal2010,Liu+etal2012,Li+etal2011,Hanson+etal2011}.
As \eg in \cite{Hanson+etal2011}, Mie scattering theory in the Rayleigh limit of small particles is used as an argument to support
the (negative) experimental conclusions made in \cite{Li+etal2011}.

In this paper, we investigate the physical realizability of achieving a RF-heating in a colloidal GNP suspension, similar as in \cite{Sassaroli+etal2012}.
However, instead of evaluating a particular phenomenological model for a GNP suspension in a particular lossy medium, 
we study the optimal absorption that can be achieved by a physically realizable passive material immersed inside any lossy medium.
Previously, optimal absorption has been studied mainly for a lossless exterior domain, such as in \cite{Miller+etal2016} giving
geometry independent absorption bounds for the plasmonic resonances in metals. In this paper, we prove that the optimal
permittivity function yielding a maximal absorption inside a spherical suspension is a conjugate match with respect to the surrounding lossy material.
Our formulation also takes the skin effect inside the surrounding lossy material into account and we show that it is the skin effect that ultimately limits
the usefulness of the local heating. 
The analysis is based on a canonical spherical geometry giving explicit quantitative answers that can be used as an indication towards the physical realizability of the heating. 
The analysis can also be used as a framework to study and evaluate any particular phenomenological parameter models such as the 
electrophoretic particle movement which can be described by the Drude model, etc., see \eg \cite{Sassaroli+etal2012,Corr+etal2012}. 
Numerical examples are included to demonstrate the feasibility of achieving electrophoretic (plasmonic) resonances in the microwave regime
in a case when the nanoparticle suspension is immersed in a weak electrolyte solution imitating a typical biological tissue.
Here, the frequency band has been chosen around 2.6\unit{GHz} constituting a realistic example based on recently proposed hyperthermia 
measurement devices and nanoparticle setups, see \eg \cite{marquez2013hyperthermia}. It should also be noted that, in practice,
the adequate frequency range will be linked to the size and concentration of the nanoparticles, 
where 5\unit{nm} is the maximum size of the GNPs to allow the particles to pass out of the kidney \cite{Callaghan+etal2010}.

\section{Optimal absorption in gold nanoparticle suspensions}

\subsection{Notation and conventions}
The following notation and conventions will be used below.
The Maxwell's equations \cite{Jackson1999} for the electric and magnetic fields $\bm{E}$ and $\bm{H}$ are considered based on SI-units 
and with time convention $\eu^{-\iu\omega t}$ for time harmonic fields.
Let $\mu_0$, $\epsilon_0$, $\eta_0$ and ${\rm c}_0$ denote the permeability, the permittivity, the wave impedance and
the speed of light in vacuum, respectively, and where $\eta_0=\sqrt{\mu_0/\epsilon_0}$ and ${\rm c}_0=1/\sqrt{\mu_0\epsilon_0}$.
The wavenumber of vacuum is given by $k_0=\omega\sqrt{\mu_0\epsilon_0}$, where $\omega=2\pi f$ is the angular frequency and $f$ the frequency.
The spherical coordinates are denoted by $(r,\theta,\phi)$, the corresponding unit vectors $(\hat{\bm{r}},\hat{\bm{\theta}},\hat{\bm{\phi}})$,
and the radius vector $\bm{r}=r\hat{\bm{r}}$.  The cartesian unit vectors are denoted $(\hat{\bm{x}},\hat{\bm{y}},\hat{\bm{z}})$.
Finally, the real and imaginary part and the complex conjugate of a complex number $\zeta$ are denoted $\Re\!\left\{\zeta\right\}$, $\Im\!\left\{\zeta\right\}$
and $\zeta^*$, respectively.

\subsection{Canonical problem setup}
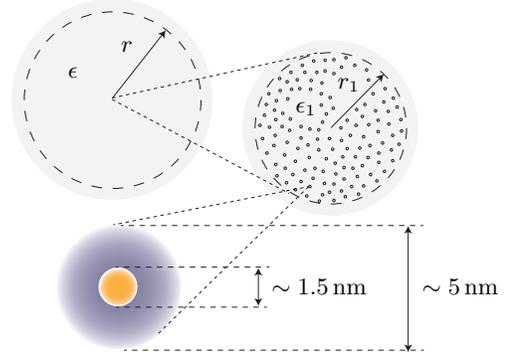
\begin{figure}[htb]
\begin{picture}(50,150)
\put(40,10){\makebox(150,140){\includegraphics[width=6cm]{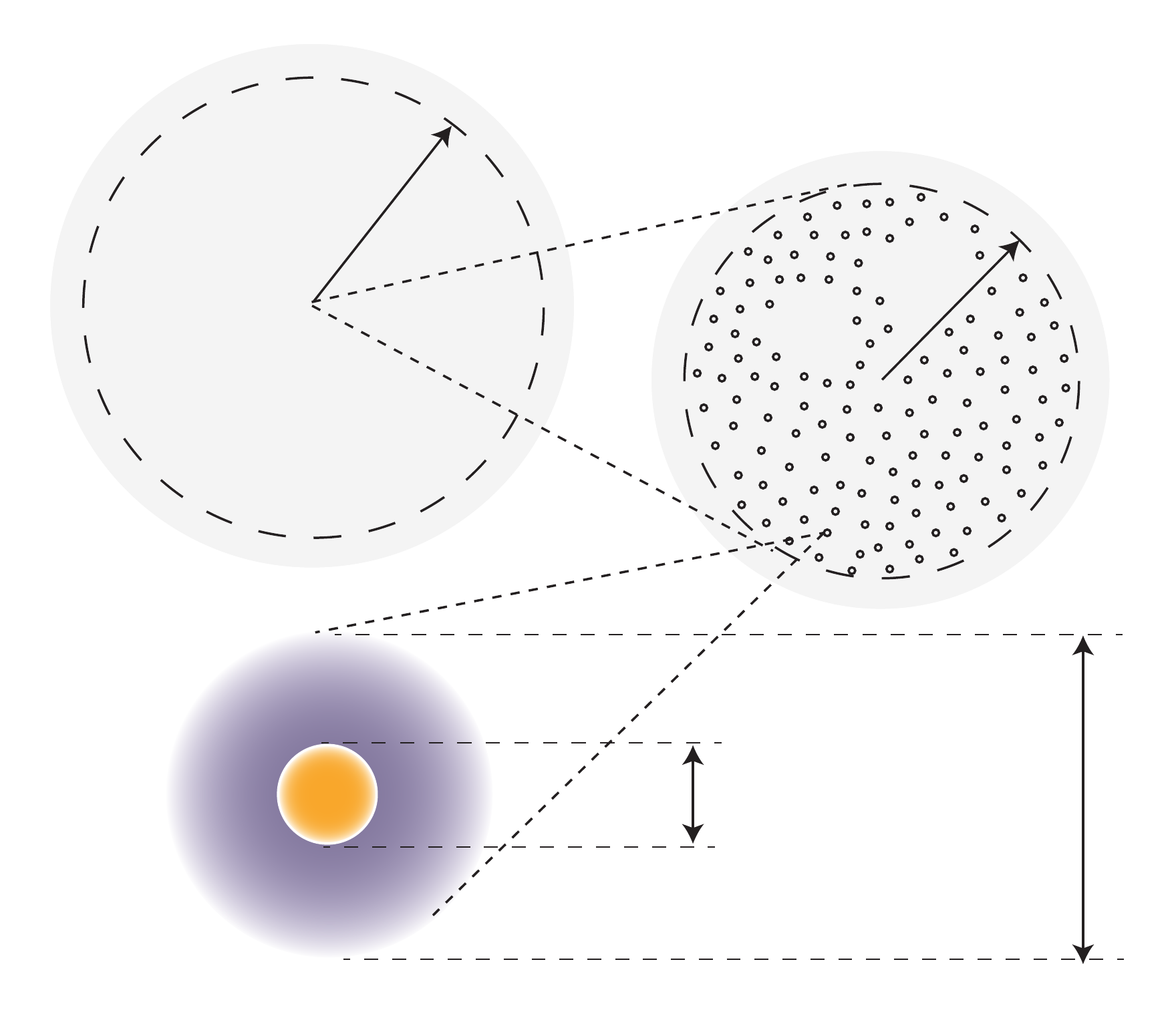}}} 
\put(157,113){ \small  $r_1$}
\put(75,127){ \small  $r$}
\put(141,104){ \small  $\epsilon_1$}
\put(55,117){ \small  $\epsilon$}
\put(132,35){ \small  $\sim 1.5$\unit{nm}}
\put(189,35){ \small  $\sim 5$\unit{nm}}
\end{picture}
\caption{Canonical problem setup based on spherical geometry. The figure also illustrates some typical dimensions for gold nanoparticles 
coated with glutathione ligands \cite{marquez2013hyperthermia,Callaghan+etal2010}.}
\label{fig:RFsetup4pdf}
\end{figure}

A simple canonical problem setup based on spherical geometry is considered as depicted in \Fref{fig:RFsetup4pdf}.
Here, $r_1$ is the radius of the spherical suspension of gold nanoparticles and $r$ the radius of a reference surface inside the surrounding medium
where the skin effect will be defined.  
The spherical suspension as well as the surrounding medium are assumed to be non-magnetic,
homogeneous and isotropic media with relative permittivity $\epsilon_1$ and $\epsilon$
and wavenumbers $k_1=k_0\sqrt{\epsilon_1}$ and $k=k_0\sqrt{\epsilon}$, respectively.

The following examples of approximative dispersion models 
will be considered here.
The surrounding medium is assumed to consist of a weak electrolyte solution with relative permittivity 
\begin{eqnarray}\label{eq:epsilondef}
\epsilon(\omega)=\epsilon_{\infty}+\frac{\epsilon_{\rm s}-\epsilon_{\infty}}{1-\iu\omega\tau}+\iu\frac{\sigma}{\omega\epsilon_0},
\end{eqnarray}
where $\epsilon_{\infty}$, $\epsilon_{\rm s}$ and $\tau$ are the high frequency permittivity,  the static permittivity and the dipole relaxation time in the 
corresponding Debye model for water, respectively, and $\sigma$ the conductivity of the saline solution. 
The suspension of gold nanoparticles is furthermore assumed to consist of a homogeneous solution of charged nanoparticles
where the electric current is governed by an electrophoretic particle acceleration mechanism \cite{Collins+etal2014,Sassaroli+etal2012}.
Hence, the effective permittivity of the nanoparticle suspension is modeled here as
\begin{eqnarray}\label{eq:epsilon1def}
\epsilon_1(\omega)=\epsilon(\omega)+\iu\frac{\sigma_1}{\omega\epsilon_0}\frac{1}{1-\iu\omega\tau_1},
\end{eqnarray}
where $\sigma_1$ is the static conductivity and $\tau_1$ the relaxation time in the corresponding Drude model.
In an approximate phenomenological description of the electrophoretic mechanism these parameters can be interpreted as
$\sigma_1={\cal N}q^2/\beta$ and $\tau_1=m/\beta$, where ${\cal N}$ is the number of particles per unit volume, 
$q$ the particle charge, $\beta$ the friction constant of the host medium and $m$ the particle mass, \cf \cite{Sassaroli+etal2012}.


\subsection{Optimal near field and the skin effect}
Consider an expansion of the electromagnetic field in terms of vector spherical waves as outlined in \ref{sect:sphericaldef}.
It is assumed that all sources are placed outside the surface of radius $r$ as depicted in \Fref{fig:RFsetup4pdf}.
It is furthermore assumed that the scattering from the nanoparticle suspension is weak and does not interact with any of the possible material obstacles outside
the radius $r$. Hence, the incident and the scattered fields for $r>r_1$ can be represented by regular and outgoing waves with multipole coeffcients $a_{\tau ml}$ 
and $b_{\tau ml}$, respectively, and the interior field for $r<r_1$ by regular waves with multipole coeffcients $a_{\tau ml}^{(1)}$.
The transition matrices give the scattering $b_{\tau ml}=t_{\tau l}a_{\tau ml}$ and the absorption $a_{\tau ml}^{(1)}=r_{\tau l}a_{\tau ml}$,
where $t_{\tau l}$ and $r_{\tau l}$ are defined in  \eref{eq:btaumldef} and \eref{eq:ataumldef} in \ref{sect:sphericalT}.

The mean local heating (in \unit{W/m^3}) generated inside the nanoparticle suspension of radius $r_1$ is given by Poynting's theorem as
\begin{eqnarray}\label{eq:Plocr1def}
P_{\rm loc}(r_1)=\frac{3}{4\pi r_1^3}\int_{V_{r_1}}\frac{1}{2}\omega\epsilon_0\Im\{\epsilon_1\}\left| \bm{E}(\bm{r}) \right|^2{\rm d} v \\
\quad =\frac{3\omega\epsilon_0\Im\{\epsilon_1\}}{8\pi r_1^3}\sum_{l=1}^{\infty}\sum_{m=-l}^{l}\sum_{\tau=1}^2  W_{\tau l}(k_1,r_1)\left| a_{\tau ml}^{(1)}\right|^2 \nonumber \\
\quad =\frac{3\omega\epsilon_0\Im\{\epsilon_1\}}{8\pi r_1^3}\sum_{l=1}^{\infty}\sum_{m=-l}^{l}\sum_{\tau=1}^2  W_{\tau l}(k_1,r_1)\left| r_{\tau l} \right|^2\left| a_{\tau ml}\right|^2, \nonumber
\end{eqnarray}
where the orthogonality of the regular vector spherical waves ${\bf v}_{\tau ml}(k_1\bm{r})$ have been used,
$V_{r_1}$ denotes the spherical volume of radius $r_1$ and where 
$W_{\tau l}(k_1,r_1)=\int_{V_{r_1}}\left|{\bf v}_{\tau ml}(k_1\bm{r})\right|^2{\rm d} v$, \cf \eref{eq:vorthogonal2} and \eref{eq:Wtauldef}. 
In \eref{eq:Plocr1def}, the relation $a_{\tau ml}^{(1)}=r_{\tau l}a_{\tau ml}$ has also been inserted.
It is shown in \ref{sect:Lommel} and \ref{sect:sphericalorth} that
\begin{eqnarray}
W_{1l}(k_1,r_1)=\frac{r_1^2 \Im\!\left\{k_1{\rm j}_{l+1}(k_1r_1) {\rm j}_{l}^*(k_1r_1) \right\}}{\Im\!\left\{k_1^2\right\}}, \\
W_{2 l}(k_1,r_1)=\frac{1}{2l+1}\left((l+1)W_{1,l-1}(k_1,r_1) \right. \\
\qquad \left. +lW_{1,l+1}(k_1,r_1) \right), \nonumber
\end{eqnarray}
where ${\rm j}_l(\cdot)$ are the spherical Bessel functions of order $l$, \cf \eref{eq:Wtauldef}, \eref{eq:W1ldef1} and \eref{eq:W2ldef}.

As a quantitative measure of the skin effect,
the mean background heating (in \unit{W/m^3}) at radius $r$ in the surrounding medium is defined by
\begin{eqnarray}\label{eq:Pbdef}
P_{\rm b}(r)=\displaystyle \frac{1}{4\pi r^2}\int_{S_r}\frac{1}{2}\omega\epsilon_0\Im\{\epsilon\}\left| \bm{E}(\bm{r}) \right|^2{\rm d} S \\
\quad =\frac{1}{8\pi}\omega\epsilon_0\Im\{\epsilon\}\int_{\Omega_0}\left| \bm{E}(\bm{r}) \right|^2{\rm d} \Omega,
\end{eqnarray}
where $S_r$ denotes the spherical boundary of radius $r$,  $\Omega_0$ the unit sphere and ${\rm d} S=r^2{\rm d}\Omega$ 
with ${\rm d}\Omega$ the differential solid angle. 
By exploiting the orthogonality of the vector spherical waves \eref{eq:vorthogonal1},
the mean background loss can now be expressed as
\begin{equation}\label{eq:Pbdef2}
P_{\rm b}(r)=\frac{1}{8\pi}\omega\epsilon_0\Im\{\epsilon\}
\sum_{l=1}^{\infty}\sum_{m=-l}^{l}\sum_{\tau=1}^2  S_{\tau l}(k,r)\left| a_{\tau ml}\right|^2,
\end{equation}
where $S_{\tau l}(k,r)=\int_{\Omega_0}|{\bf v}_{\tau ml}(k\bm{r})|^2{\rm d}\Omega$, see also \eref{eq:Stauldef}.

The optimal near field is now defined by the maximization of the relative heating coefficient
 $F=P_{\rm loc}(r_1)/P_{\rm b}(r)$ at some radius $r$. 
This power ratio is a generalized Rayleigh quotient and hence the problem is equivalent to finding the maximum eigenvalue
in the corresponding (diagonal) generalized eigenvalue problem as follows
\begin{equation}\label{eq:Plocr1Pbr2}
\displaystyle \max_{\left|a_{\tau ml}\right|^2} \frac{P_{\rm loc}(r_1)}{P_{\rm b}(r)} 
= \frac{3}{r_1^3}\frac{\Im\{\epsilon_1\}}{\Im\{\epsilon\}} \max_{\tau, l} \frac{W_{\tau l}(k_1,r_1)\left| r_{\tau l} \right|^2}{S_{\tau l}(k,r)}.
\end{equation}
With the relatively low frequencies and small geometrical dimensions of interest in this paper, the optimal near field excitation
is generally obtained with an electric dipole field ($\tau=2$ and $l=1$ in \eref{eq:Plocr1Pbr2}) yielding the relative heating coefficient
\begin{eqnarray}\label{eq:relheatingdef1}
F(\epsilon_1)=\frac{3}{r_1^3}\frac{\Im\{\epsilon_1\}}{\Im\{\epsilon\}}  \frac{W_{21}(k_1,r_1)\left| r_{21} \right|^2}{S_{21}(k,r)},
\end{eqnarray}
where $F(\epsilon_1)$ is considered to be a function of the complex-valued permittivity $\epsilon_1$ (or frequency $\omega$) when all other parameters are fixed.

\subsection{Asymptotic analysis}
Based on the asymptotic expansions of the spherical Bessel and Hankel functions for small arguments \cite{Olver+etal2010}
${\rm j}_0(x)\sim 1$, ${\rm j}_1(x)\sim \frac{1}{3}x$, ${\rm j}_1^{\prime}(x)\sim \frac{1}{3}$, ${\rm j}_2(x)\sim \frac{1}{15}x^2$, ${\rm j}_3(x)\sim \frac{1}{105}x^3$, 
${\rm h}_1^{(1)}(x)\sim -\frac{\iu}{x^2}$ and ${\rm h}_1^{(1)\prime}(x)\sim \frac{2\iu}{x^3}$,
it is found that 
\begin{eqnarray}
W_{21}\sim \frac{2}{9}r_1^3, \label{eq:W21a}\\
r_{21}\sim \frac{3\epsilon}{\epsilon_1+2\epsilon}, \label{eq:r21a}
\end{eqnarray}
in the asymptotic limit of small $r_1$. Hence, for small $r_1$ the asymptotic relative heating coefficient can be expressed as
\begin{eqnarray}\label{eq:relheatingdef2}
F^{\rm a}(\epsilon_1)=\frac{F^{\rm a}_{\rm num}(\epsilon_1)}{S_{21}(k,r)},
\end{eqnarray}
where the numerator is given by
\begin{eqnarray}\label{eq:relheatingnum}
F^{\rm a}_{\rm num}(\epsilon_1)=6\frac{|\epsilon|^2}{\Im\{\epsilon\}}\frac{\Im\{\epsilon_1\}}{|\epsilon_1-\epsilon_{\rm 1o}^*|^2},
\end{eqnarray}
and where the quantity $\epsilon_{\rm 1o}$ is defined by $\epsilon_{\rm 1o}^*=-2\epsilon$. 
Note that $F^{\rm a}(\epsilon_1)$ is independent of the radius $r_1$.

\subsection{Optimal absorption and the conjugate match}
Although it is not easy to prove in general that $F(\epsilon_1)$ defined in \eref{eq:relheatingdef1} is a convex function of $\epsilon_1$ for $\Im\epsilon_1>0$, it is straightforward to
show that $\epsilon_{\rm 1o}=-2\epsilon^*$ is a local maximum of $F^{\rm a}_{\rm num}(\epsilon_1)$ defined in \eref{eq:relheatingnum}.
Hence, let $F_0(\epsilon_1,\epsilon_1^*)=\frac{\Im\epsilon_1}{|\epsilon_1-\epsilon_{\rm 1o}^*|^2}$ and consider the following Taylor expansion
in a neighbourhood of $\epsilon_{\rm 1o}$
\begin{eqnarray}
F_0(\epsilon_1,\epsilon_1^*)=-\iu\frac{\epsilon_1-\epsilon_1^*}{2}\frac{1}{\epsilon_1-\epsilon_{\rm 1o}^*}\frac{1}{\epsilon_1^*-\epsilon_{\rm 1o}} \\
=F_0(\epsilon_{\rm 1o},\epsilon_{\rm 1o}^*)+\frac{\partial F_0}{\partial \epsilon_1}(\epsilon_1-\epsilon_{\rm 1o})
+\frac{\partial F_0}{\partial \epsilon_1^*}(\epsilon_1^*-\epsilon_{\rm 1o}^*) \nonumber \\
+\frac{1}{2}\frac{\partial^2 F_0}{\partial \epsilon_1^2}(\epsilon_1-\epsilon_{\rm 1o})^2
+\frac{1}{2}\frac{\partial^2 F_0}{\partial \epsilon_1^{*2}}(\epsilon_1^*-\epsilon_{\rm 1o}^*)^2 \nonumber \\
+\frac{\partial^2 F_0}{\partial \epsilon_1\partial\epsilon_1^*}(\epsilon_1-\epsilon_{\rm 1o})(\epsilon_1^*-\epsilon_{\rm 1o}^*)+\cdots, \nonumber
\end{eqnarray}
where the complex derivatives $\frac{\partial}{\partial \epsilon_1}$ and $\frac{\partial}{\partial \epsilon_1^*}$ are defined as in \cite{Greene+Krantz1997}.
It is straightforward to show that
\begin{eqnarray}
\frac{\partial F_0}{\partial \epsilon_1}=\frac{\partial F_0}{\partial \epsilon_1^*}=\frac{\partial^2 F_0}{\partial \epsilon_1^2}=\frac{\partial^2 F_0}{\partial \epsilon_1^{*2}}=0
\end{eqnarray}
at $\epsilon_1=\epsilon_{\rm 1o}$ and
\begin{eqnarray}
\frac{\partial^2 F_0}{\partial \epsilon_1\partial\epsilon_1^*}=-\frac{\Im\{\epsilon_{\rm 1o}\}}{|\epsilon_1-\epsilon_{\rm 1o}^*|^4},
\end{eqnarray}
which is negative definite when $\Im\{\epsilon_{\rm 1o}\}>0$. It is noted that the passivity of the external material with $\Im\{\epsilon\}>0$ guarantees that
$\epsilon_{\rm 1o}=-2\epsilon^*$ has a positive imaginary part. Hence, the conjugate match\footnote{Note that the factor 2 in $\epsilon_{\rm 1o}=-2\epsilon^*$
is a form factor associated with the spherical geometry, see for instance [19, p. 145].} $\epsilon_{\rm 1o}=-2\epsilon^*$ yields
a maximum of the relative heating function $F^{\rm a}(\epsilon_1)$ in \eref{eq:relheatingdef2} and an optimal absorption inside the suspension domain with radius $r_1$.

\subsection{Relation to general Mie theory}
The expression for the absorption cross section $C_{\rm abs}$ of a small homogeneous dielectric sphere in a lossy medium 
is obtained from general Mie theory as
\begin{eqnarray}\label{eq:Cabs}
C_{\rm abs}=C_{\rm inc}-C_{\rm sca}+C_{\rm ext},
\end{eqnarray}
where the scattering cross section $C_{\rm sca}$, the extinction cross section $C_{\rm ext}$ and the compensation term $C_{\rm inc}$
(power absorbed from the incident plane wave in the surrounding medium) are given by
\begin{eqnarray}
C_{\rm sca}=\frac{16\pi}{3}k_0r_1^3\Im\{\sqrt{\epsilon}\}\left|\frac{\epsilon_1-\epsilon}{\epsilon_1+2\epsilon} \right|^2, \label{eq:Csca} \\
C_{\rm ext}=6\pi k_0r_1^3\left[\frac{4}{9}\Re\left\{\frac{\epsilon_1-\epsilon}{\epsilon_1+2\epsilon}\right\}\Im\{\sqrt{\epsilon}\} \right. \nonumber \\
\left. +\frac{2}{3}\Im\left\{\frac{\epsilon_1-\epsilon}{\epsilon_1+2\epsilon}\right\}
\left(\Re\{\sqrt{\epsilon}\}-\frac{(\Im\{\sqrt{\epsilon}\})^2}{\Re\{\sqrt{\epsilon}\}} \right) \right], \label{eq:Cext} \\
C_{\rm inc}=\frac{8\pi}{3}k_0r_1^3\Im\{\sqrt{\epsilon}\},\label{eq:Ci}
\end{eqnarray}
see \eg \cite{Sassaroli+etal2012,Bohren+Huffman1983,Bohren+Gilra1979,Chylek1977}.
By algebraic manipulation of \eref{eq:Cabs} through \eref{eq:Ci} (or by performing the derivation from first principles),
it can be shown that the absorption cross section is given by 
\begin{eqnarray}\label{eq:Cabs2}
C_{\rm abs}=12\pi k_0r_1^3 \frac{|\epsilon|^2}{\Re\{\sqrt{\epsilon}\}}\frac{\Im\{\epsilon_1\}}{\left|\epsilon_1+2\epsilon \right|^2},
\end{eqnarray}
which has a striking resemblance to the heating coefficent given by \eref{eq:relheatingnum}. Hence, the resulting expression \eref{eq:Cabs2} immediately verifies
that the conjugate match $\epsilon_{\rm 1o}=-2\epsilon^*$  is also optimal with respect to the absorption cross section 
associated with a small dielectric sphere in a lossy medium in accordance with the Mie theory.

\subsection{Narrowband realizability of the conjugate match: Tuning the Drude model}
Now that we have shown that the conjugate match $\epsilon_{\rm 1o}=-2\epsilon^*$ yields the optimal absorption at any given frequency $\omega$,
the next step is to consider the realizability of an interior permittivity function $\epsilon_1(\omega)$ that can achieve this property. With a ``normal'' surrounding medium having
a permittivity function $\epsilon$ with $\Re\{\epsilon\}>0$, this requires a permittivity function for the interior region with a negative real part $\Re\{\epsilon_1\}<0$,
a material property that sometimes is associated with a {\em metamaterial} and which may have severe limitations on the bandwidth capabilities, \cf \cite{Gustafsson+Sjoberg2010a}. 

Given that the approximate electrophoretic particle acceleration mechanism described above is valid, it
is straightforward to ``tune'' the corresponding Drude model in \eref{eq:epsilon1def} to resonance at the desired frequency $\omega_{\rm d}$ by solving the equation 
\begin{eqnarray}\label{eq:tuningDrude1}
\epsilon_1(\omega_{\rm d})=\epsilon(\omega_{\rm d})+\iu\frac{\sigma_1}{\omega_{\rm d}\epsilon_0}\frac{1}{1-\iu\omega_{\rm d}\tau_1}=\epsilon_{\rm 1o}(\omega_{\rm d}),
\end{eqnarray}
yielding the following tuned parameters
\begin{eqnarray}\label{eq:tuningDrude2a}
\tau_1=\frac{1}{\omega_{\rm d}}\frac{\Re\epsilon(\omega_{\rm d})-\Re\epsilon_{\rm 1o}(\omega_{\rm d})}{\Im\epsilon_{\rm 1o}(\omega_{\rm d})-\Im\epsilon(\omega_{\rm d})}, \\
\sigma_1=\epsilon_0\left(\Re\epsilon(\omega_{\rm d})-\Re\epsilon_{\rm 1o}(\omega_{\rm d}) \right)\frac{1+\omega_{\rm d}^2\tau_1^2}{\tau_1}.\label{eq:tuningDrude2b}
\end{eqnarray}
It should be noted that the inclusion of the term $\epsilon(\omega)$ in \eref{eq:epsilon1def} and \eref{eq:tuningDrude1} above is ``ad hoc'',
and can be replaced for any other model as long as the term $\Im\epsilon_{\rm 1o}(\omega_{\rm d})-\Im\epsilon(\omega_{\rm d})>0$ in \eref{eq:tuningDrude2a} above.

\subsection{Broadband realizability of the conjugate match: Optimal dispersion modeling}
To study the optimal capabilities of a passive material with permittivity $\epsilon_1(\omega)$ to approximate 
the desired conjugate match $\epsilon_{\rm 1o}(\omega)=-2\epsilon^*(\omega)$ over a given bandwidth, 
a convex optimization approach is employed as follows, see also \cite{Nordebo+etal2014b}.
The permittivity function of any passive material $\epsilon_1(\omega)$ corresponds to a symmetric Herglotz function
$h_1(\omega)=\omega\epsilon_1(\omega)$, where $\omega\in\C_+=\{\omega\in\C|\Im\omega>0\}$, 
see \eg \cite{Zemanian1965,Kac+Krein1974,Akhiezer1965,Nussenzveig1972,Gustafsson+Sjoberg2010a,Bernland+etal2011b}.
Symmetric Herglotz functions are analytic functions with symmetry $h_1(\omega)=-h_1^*(-\omega^*)$ 
mapping the upper half-plane into itself and which allow for an integral representation
based on positive measures. On the real line this integral representation becomes a Cauchy-principal value integral (Hilbert transform) acting on
the regular part of the measure plus the contribution from possible point masses. Here, the limiting Herglotz function for $\omega\in \R$
is represented by
\begin{eqnarray}
h_1(\omega)=\omega\epsilon_{1\infty}+\frac{1}{\pi}\dashint_{-\infty}^{\infty}\frac{1}{\xi-\omega}\Im h_{\rm 1r}(\xi){\rm d}\xi +x_0\frac{1}{-\omega} \nonumber \\
+\iu\left(\Im h_{\rm 1r}(\omega)+\pi x_0\delta(\omega)\right),\label{eq:h1repr1}
\end{eqnarray}
where $\epsilon_{1\infty}$ is the high frequency permittivity,
$\Im h_{\rm 1r}(\xi)$ the regular part of the positive symmetric measure and where a point mass with amplitude $x_0\geq 0$ at $\omega=0$ has been included.
Obviously, the point mass at $\omega=0$ will be very efficient in generating a negative real part of $h_1(\omega)$ at any particular desired frequency $\omega_{\rm d}$.

Next, a finite optimization domain $\Omega_{\rm r}$  (support of the regular measure) is defined with $0\notin\Omega_{\rm r}$ and where
the positive symmetric measure is approximated by using 
\begin{eqnarray}\label{eq:h1repr2}
\Im h_{\rm 1r}(\omega)=\sum_{n=1}^{N}x_n\left[p_n(\omega) +p_n(-\omega) \right],
\end{eqnarray}
where $p_n(\omega)$ are triangular basis functions defined on $\Omega_{\rm r}$ 
and $x_n\geq 0$ the corresponding optimization variables for $n=1,\ldots,N$.
Further, a finite approximation domain $\Omega\subset\Omega_{\rm r}$ is defined
and the corresponding real part $\Re h_{\rm 1r}(\omega)$ on $\Omega$ is given by
\begin{eqnarray}
\Re h_{\rm 1r}(\omega)=
\sum_{n=1}^{N}x_n\left[\hat{p}_n(\omega) -\hat{p}_n(-\omega) \right],  \label{eq:h1repr3}
\end{eqnarray}
where $\hat{p}_n(\omega)=\displaystyle\frac{1}{\pi}\dashint_{-\infty}^{\infty}\frac{1}{\xi-\omega}p_n(\xi){\rm d}\xi$ 
is the (negative) Hilbert transform of the triangular pulse function $p_n(\omega)$, \cf \cite{King2009}.
The discrete representation \eref{eq:h1repr1} through \eref{eq:h1repr3} can now be used to formulate
the following convex optimization problem
\begin{eqnarray}\label{eq:cvxdef}
\begin{array}{llll}
	& \minimize & & \| h_1(\omega)-f(\omega) \|_\Omega  \\    
	& \subto & &  x_n \geq 0, \\  
	&       &  & \epsilon_{1\infty} \geq 1,  
\end{array}
\end{eqnarray}
where the target function is $f(\omega)=\omega\epsilon_{\rm 1o}(\omega)$ and $\|\cdot\|_\Omega$ a suitable norm defined on $\Omega$,
see also \cite{Nordebo+etal2014b}. Here, $x_n$ with $n=0,1,\ldots,N$ and $\epsilon_{1\infty}$ are the positive optimization variables.
The optimization problem \eref{eq:cvxdef} can be solved efficiently using the CVX Matlab software for disciplined convex programming \cite{Grant+Boyd2012}.
Typically, $\Omega$ consists of a set of frequency points sampled around the desired center frequency $\omega_{\rm d}$. 
If $\omega_{\rm d}$ is large (such as in the \unit{GHz} range), it is usually necessary to employ scaled dimensionless variables $x_n/\omega_{\rm d}$
for $n=1,\dots,N$ and $x_0/\omega_{\rm d}^2$ to maintain numerical stability.

\section{Numerical examples}
To illustrate the theory above the following numerical  example is considered. The surrounding medium is assumed to consist of a weak electrolyte
solution with relative permittivity given by \eref{eq:epsilondef} and where $\epsilon_{\infty}=5.27$, $\epsilon_{\rm s}=80$ and $\tau=10^{-11}$\unit{s} 
are the Debye parameters for water and $\sigma\in\{0.1,1,10\}$\unit{S/m} the conductivity parameters of the saline solution.
The radii of the reference surface and the spherical suspension are $r=5$\unit{cm} and $r_1=1$\unit{\micro m}, respectively.
In all of the numerical results below, the optimality of using electric dipole excitation ($\tau=2$, $l=1$) has been verified by
evaluating \eref{eq:Plocr1Pbr2} with $\tau=1,2$ and $l=1,2,3,\ldots$, etc.


\begin{figure}[htb]
\begin{picture}(50,140)
\put(90,0){\makebox(50,130){\includegraphics[width=7.5cm]{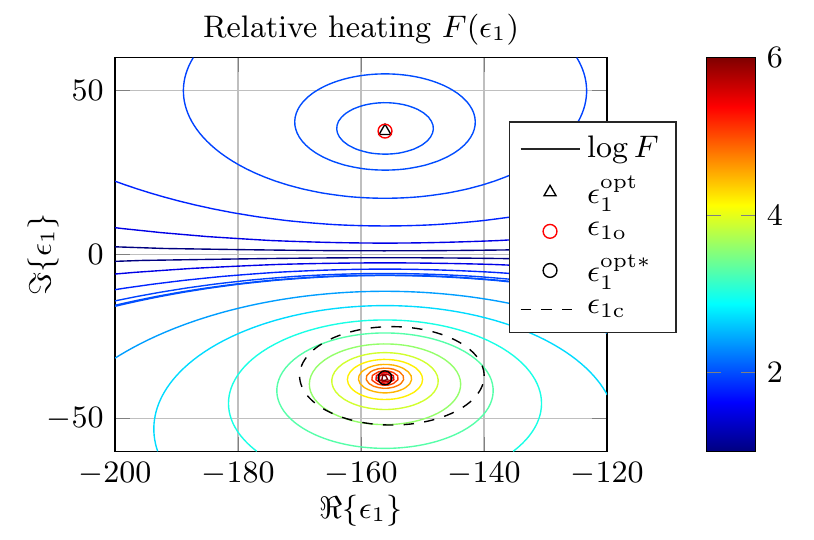}}} 
\end{picture}
\caption{Contour plot of the relative heating $F(\epsilon_1)$ as a function of the complex permitivity $\epsilon_1$
at frequency $f_{\rm d}=2.6\unit{GHz}$. The surrounding medium is a saline solution with conductivity $\sigma=1$\unit{S/m}.}
\label{fig:matfig143}
\end{figure}

In \Fref{fig:matfig143} is shown a contour plot of the relative heating coefficient $F(\epsilon_1)$ given by the rigorous expression \eref{eq:relheatingdef1}.
The surrounding medium is a saline solution with conductivity $\sigma=1$\unit{S/m} and the desired center frequency is $f_{\rm d}=2.6\unit{GHz}$.
The plot clearly illustrates the concavity of the target function $F(\epsilon_1)$ in the upper half-plane $\Im\epsilon_1>0$.
Here, $\epsilon_1^{\rm opt *}$ is obtained from a numerical residue calculation in the lower half-plane
as $\epsilon_1^{\rm opt *}=(\oint_{\epsilon_{\rm 1c}}a^{-1}(\epsilon_1)\epsilon_1 {\rm d}\epsilon_1)/(\oint_{\epsilon_{\rm 1c}}a^{-1}(\epsilon_1){\rm d}\epsilon_1)$,
where $a(\epsilon_1)$ is the analytic denominator (determinant) associated with $r_{21}$ defined in \eref{eq:ataumldef} and $\epsilon_{\rm 1c}$ an arbitrary contour
circumscribing the pole (zero) location of $r_{21}$ ($a(\epsilon_1)$). Further, $\epsilon_{\rm 1o}$ denotes the conjugate match $\epsilon_{\rm 1o}=-2\epsilon^*$
which is the asymptotic solution that yields a maximal absorption for small $r_1$. In this numerical example
$\epsilon_1^{\rm opt }=-156.12 + \iu 37.609$ and $\epsilon_{\rm 1o} =-156.11 + \iu 37.609$,
demonstrating the high accuracy of the asymptotic solution.

\subsection{Tuning the Drude model}\label{sect:extuningDrude}
\begin{figure}[htb]
\begin{picture}(50,140)
\put(90,0){\makebox(50,130){\includegraphics[width=7.5cm]{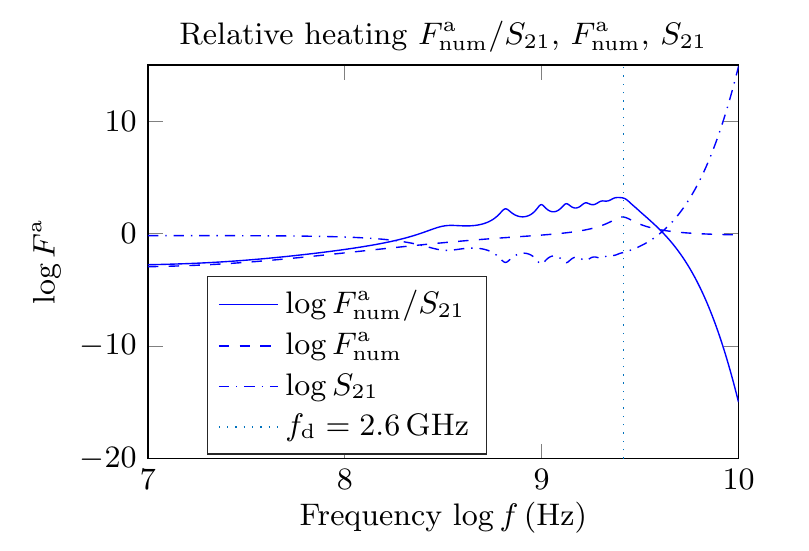}}} 
\end{picture}
\caption{Relative heating $F^{\rm a}=F_{\rm num}^{\rm a}/S_{21}$ and its constituents $F_{\rm num}^{\rm a}$ and $S_{21}$ plotted as functions of frequency. 
Here, $\epsilon_1(\omega)$ is given by the Drude model  tuned to resonance at $f_{\rm d}=2.6\unit{GHz}$ and the surrounding medium
is a saline solution with conductivity $\sigma=0.1$\unit{S/m}.}
\label{fig:matfig42}
\end{figure}

In \Fref{fig:matfig42} is shown the constituents of $F^{\rm a}=F_{\rm num}^{\rm a}/S_{21}$
given by \eref{eq:relheatingdef2} and \eref{eq:relheatingnum} plotted as functions of frequency
and where $\epsilon_1(\omega)$ is given by the Drude model \eref{eq:epsilon1def} tuned to resonance at $f_{\rm d}=2.6\unit{GHz}$ 
by using \eref{eq:tuningDrude2a} and \eref{eq:tuningDrude2b}. Here, $\sigma=0.1$\unit{S/m} and the maximum relative heating 
is $F^{\rm a}=1614$ close to the center frequency at $f_{\rm d}=2.6\unit{GHz}$.


The oscillations seen at lower frequencies are merely ``resonances'' associated with the reference surface at $r=5$\unit{cm}
and are visible also in the function $S_{21}$. The increase in $S_{21}$ at higher frequencies is a manifestation of the skin effect in the surrounding medium
which will ultimately limit the potential of relative heating.


\begin{figure}[htb]
\begin{picture}(50,140)
\put(90,0){\makebox(50,130){\includegraphics[width=7.5cm]{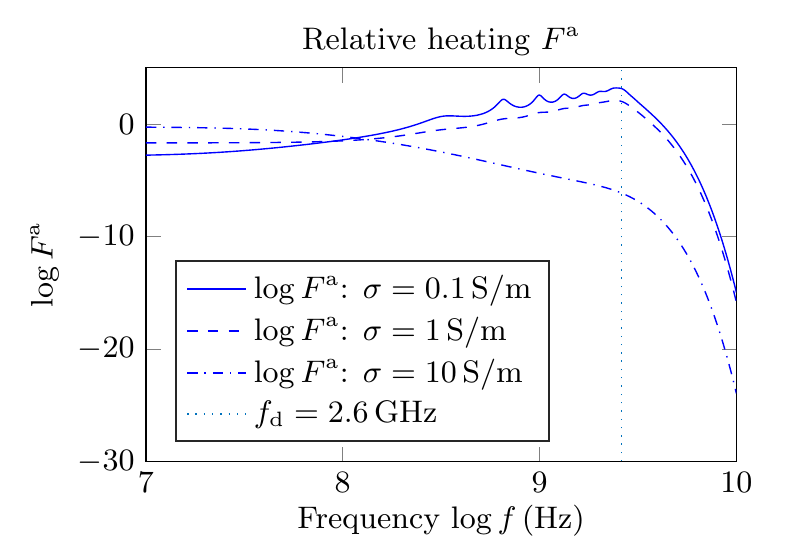}}} 
\end{picture}
\caption{Relative heating $F^{\rm a}$ as a function of frequency. 
Here, $\epsilon_1(\omega)$ is given by the Drude model  tuned to resonance at $f_{\rm d}=2.6\unit{GHz}$ and the surrounding medium
is a saline solution with conductivity  $\sigma\in\{0.1,1,10\}$\unit{S/m}.}
\label{fig:matfig4}
\end{figure}


\begin{figure}[htb]
\begin{picture}(50,140)
\put(90,0){\makebox(50,130){\includegraphics[width=7.5cm]{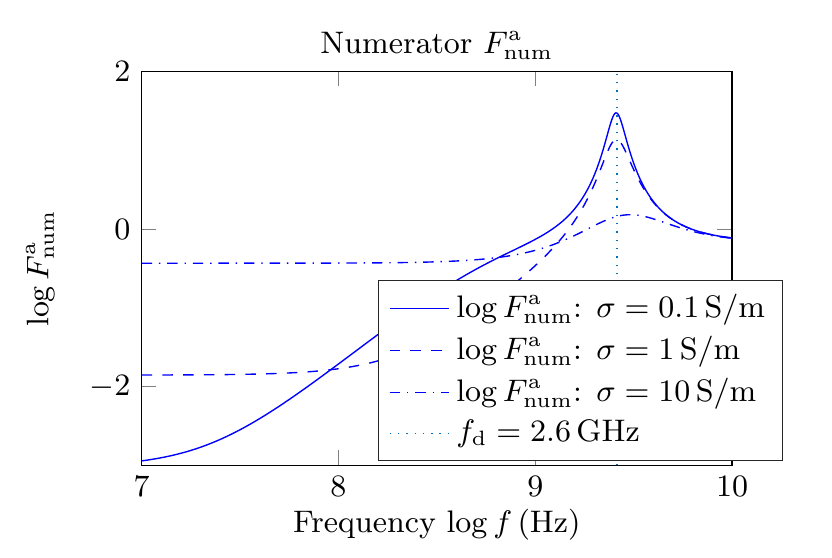}}} 
\end{picture}
\caption{Numerator $\log F_{\rm num}^{\rm a}$ as a function of frequency. 
Here, $\epsilon_1(\omega)$ is given by the Drude model  tuned to resonance at $f_{\rm d}=2.6\unit{GHz}$ and the surrounding medium
is a saline solution with conductivity  $\sigma\in\{0.1,1,10\}$\unit{S/m}.}
\label{fig:matfig5}
\end{figure}

In Figures \ref{fig:matfig4} and \ref{fig:matfig5} are shown the corresponding optimal relative heating $F^{\rm a}$ and numerator
$F_{\rm num}^{\rm a}$ for $\sigma\in\{0.1,1,10\}$\unit{S/m}. The corresponding optimally tuned Drude parameters are
$\tau_1\in\{11.4,7.6,1.8\}\cdot 10^{-10}$\unit{s} and $\sigma_1\in\{632,425,110\}\unit{S/m}$.
The corresponding maximal relative heating at $f_{\rm d}=2.6\unit{GHz}$ is obtained as $F^{\rm a}\in\{1614,126,10^{-6}\}$ illustrating
in this example a great potential of local heating  when $\sigma\in\{0.1,1\}$\unit{S/m}.
However, the result for $\sigma=10$\unit{S/m} implies that the surrounding medium is too lossy and the resulting skin effect will render the local heating useless.


\begin{figure}[htb]
\begin{picture}(50,130)
\put(90,0){\makebox(50,120){\includegraphics[width=7.5cm]{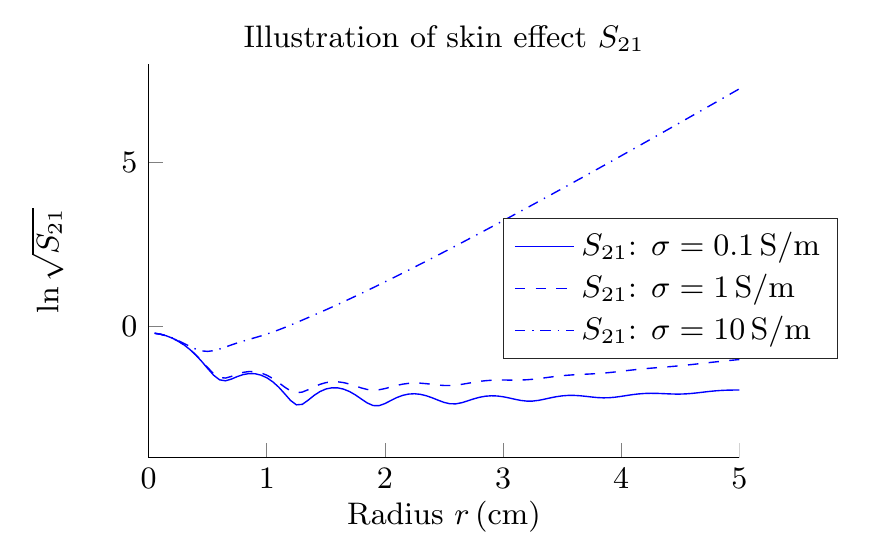}}} 
\end{picture}
\caption{Illustration of the skin effect with $\ln \sqrt{S_{21}}$ plotted as a function of radius $r$ at $f_{\rm d}=2.6$\unit{GHz} and where $\sigma\in\{0.1,1,10\}$\unit{S/m}.}
\label{fig:matfig71}
\end{figure}

The skin effect is further illustrated in \Fref{fig:matfig71} with $\ln \sqrt{S_{21}}$ plotted as a function of radius $r$ at $f_{\rm d}=2.6$\unit{GHz} and where $\sigma\in\{0.1,1,10\}$\unit{S/m}.
The skin depth can be interpreted here as the radial distance through which the amplitude $\sqrt{S_{21}}$ has decreased by a factor of $\eu^{-1}$.
Hence, the skin depth in the example case with $\sigma=10$\unit{S/m} is seen to be less than 1\unit{cm}.

\subsection{Optimal dispersion modeling}\label{sect:exoptdispmodeling}
The convex optimization formulation \eref{eq:cvxdef} is employed here
to study the optimal capabilities of a passive material with permittivity $\epsilon_1(\omega)$ to approximate the
desired conjugate match $\epsilon_{\rm 1o}(\omega)$ over a given bandwidth.
The norm is defined by
$\| h_1(\omega)-f(\omega) \|_\Omega=\sup_{\Omega}|(h_1(\omega)-f(\omega))/\omega|=\sup_{\Omega}|\epsilon_1(\omega)-\epsilon_{\rm 1o}(\omega)|$.
Here, the target function is the conjugate match $\epsilon_{\rm 1o}(\omega)=-2\epsilon^*(\omega)$, where $\epsilon(\omega)$ is 
the permittivity function of the surrounding medium given by \eref{eq:epsilondef} with Debye parameters $\epsilon_{\infty}=5.27$, $\epsilon_{\rm s}=80$ and $\tau=10^{-11}$\unit{s} 
as above and $\sigma=1$\unit{S/m}. The optimization domain for dispersion modeling (support of the regular measure) is $\Omega_{\rm r}=[2.4,2.8]$\unit{GHz} with
$N=300$ triangular basis functions and corresponding optimization variables $x_n$ for $n=1,\ldots,N$. 
Further optimization variables are the high frequency permittivity $\epsilon_{1\infty}$ and the amplitude $x_0$ of the point mass at $\omega=0$.
The finite approximation domain $\Omega$ (support of the norm) is defined by the nodal points of the approximating triangular basis functions
sampled in the interval $[2.5,2.7]$\unit{GHz}, which is centered at the desired frequency $f_{\rm d}=2.6\unit{GHz}$.


\begin{figure}[htb]
\begin{picture}(50,140)
\put(90,0){\makebox(50,130){\includegraphics[width=7.5cm]{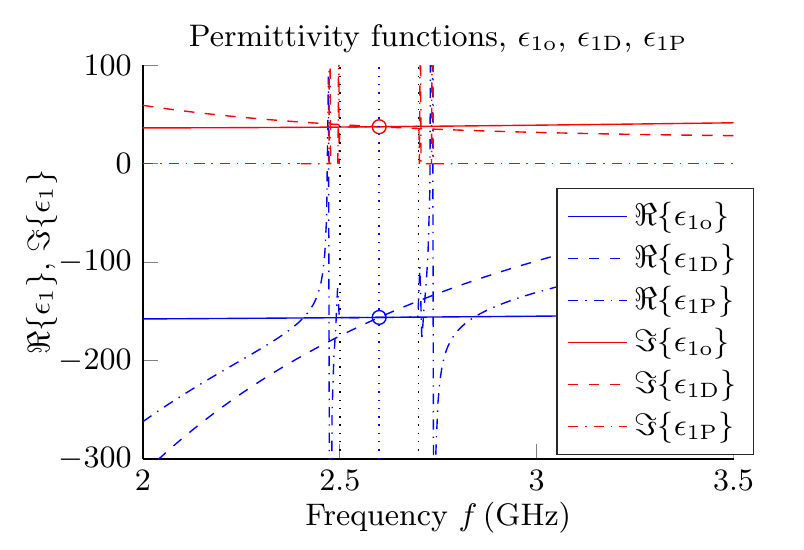}}} 
\end{picture}
\caption{Permittivity functions corresponding to the target conjugate match $\epsilon_{\rm 1o}$, the tuned Drude model $\epsilon_{\rm 1D}$ and the optimized passive model $\epsilon_{\rm 1P}$.
The surrounding medium is a saline solution with conductivity $\sigma=1$\unit{S/m} and the approximation domain is $\Omega=[2.5,2.7]$\unit{GHz}.}
\label{fig:matfig92}
\end{figure}

In \Fref{fig:matfig92} is shown the target permittivity $\epsilon_{\rm 1o}(\omega)$, 
the tuned Drude model $\epsilon_{\rm 1D}(\omega)$ according to section \ref{sect:extuningDrude} above
and the resulting optimum passive (realizable) model $\epsilon_{\rm 1P}(\omega)$, plotted as functions of frequency between 2 and 3.5\unit{GHz}.
The optimized model $\epsilon_{\rm 1P}(\omega)$ consists of a high frequency permittivity $\epsilon_{1\infty}=1$, 
a point mass at $\omega=0$ with amplitude $x_0=160.5\omega_{\rm d}^2$ and a regular measure $\Im h_{\rm 1r}(\omega)$ yielding
an impulsive response just outside and close to the approximation domain and finally a response that closely matches the
target permittivity $\epsilon_{\rm 1o}(\omega)$ inside the approximation domain. Except for the support of these responses,
$\Im h_{\rm 1r}(\omega)$ is essentially zero elsewhere and the solution is hence independent of an increase of 
the optimization domain $\Omega_{\rm r}$.


\begin{figure}[htb]
\begin{picture}(50,140)
\put(90,0){\makebox(50,130){\includegraphics[width=7.5cm]{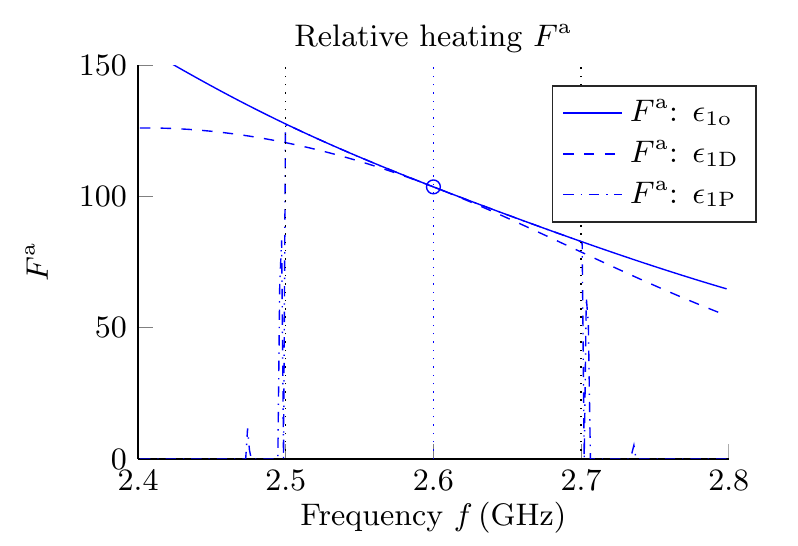}}} 
\end{picture}
\caption{Relative heating $F^{\rm a}$ corresponding to the target conjugate match $\epsilon_{\rm 1o}$, the tuned Drude model $\epsilon_{\rm 1D}$ and the optimized passive model $\epsilon_{\rm 1P}$.
The surrounding medium is a saline solution with conductivity $\sigma=1$\unit{S/m} and the approximation domain is $\Omega=[2.5,2.7]$\unit{GHz}.}
\label{fig:matfig6}
\end{figure}


\begin{figure}[htb]
\begin{picture}(50,90)
\put(90,0){\makebox(50,80){\includegraphics[width=7.5cm]{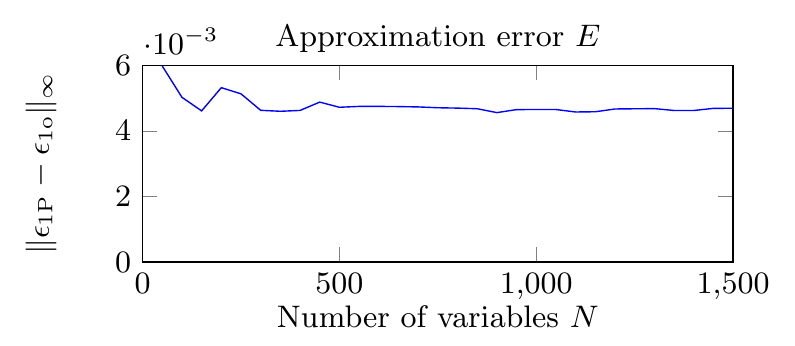}}} 
\end{picture}
\caption{Approximation error $E=\|\epsilon_{\rm 1P}-\epsilon_{\rm 1o}\|_\infty$ as a function of the number of approximating basis functions $N$.
The surrounding medium is a saline solution with conductivity $\sigma=1$\unit{S/m} and the approximation domain is $\Omega=[2.5,2.7]$\unit{GHz}.
The target permittivity is the desired conjugate match $\epsilon_{\rm 1o}(\omega)=-2\epsilon^*(\omega)$ for $\omega\in\Omega$.}
\label{fig:matfig100}
\end{figure}

In \Fref{fig:matfig92} it is further noted that $\epsilon_{\rm 1o}(\omega)$ is almost constant over the bandwidth of interest,
and it was found that the optimized result $\epsilon_{\rm 1P}(\omega)$ does
not change notably if the target function is chosen according to the constant value $-156.1 + \iu 37.61\approx\epsilon_{\rm 1o}(\omega)$ for $\omega\in\Omega$.
In \Fref{fig:matfig6} is plotted the corresponding relative heating $F^{\rm a}$ according to the target conjugate match $\epsilon_{\rm 1o}(\omega)$, the tuned Drude model
$\epsilon_{\rm 1D}(\omega)$ and the optimized model $\epsilon_{\rm 1P}(\omega)$, respectively, plotted as functions of frequency between 2.4 and 2.8\unit{GHz}.
Note that the optimal heating capability increases at lower frequencies and that the Drude model is tuned here to the desired center frequency $f_{\rm d}=2.6\unit{GHz}$.
It is seen that the optimized passive model $\epsilon_{\rm 1P}(\omega)$ closely matches the heating obtained from the target conjugate match $\epsilon_{\rm 1o}(\omega)$
when evaluated inside the approximation domain between 2.5 and 2.7\unit{GHz}.
However, there is in fact a remaining approximation error between $\epsilon_{\rm 1P}(\omega)$ and $\epsilon_{\rm 1o}(\omega)$, even though it is so small that it is not visible
in Figures \ref{fig:matfig92} and \ref{fig:matfig6}. Hence, in \Fref{fig:matfig100} is shown
the approximation error $E=\sup_{\Omega}|\epsilon_{\rm 1P}(\omega)-\epsilon_{\rm 1o}(\omega)|$ plotted as a function of the number of basis functions used $N$.
It is seen that the remaining error converges to a small non-zero value for large $N$.

 \subsection{A comment on metamaterials}
 When applying the conjugate match for a lossless background, the result is $\epsilon_{\rm 1o}=-2\epsilon$, which is a negative real number.
 A material with a negative real permittivity over a specified bandwidth is sometimes regarded a {\em metamaterial}.
In \cite{Gustafsson+Sjoberg2010a} is shown that a metamaterial  with a constant real-valued 
and negative target permittivity $\epsilon_{\rm t}<0$ over a finite frequency band $\Omega$ can only be approximated by a passive permittivity function $\epsilon_{\rm P}(\omega)$ 
within the following lower bound 
\begin{eqnarray}\label{eq:sumruleconstraint}
\quad \sup_{\omega\in\Omega}|\epsilon_{\rm P}(\omega)-\epsilon_{\rm t}|\geq \frac{(\epsilon_{\infty}-\epsilon_{\rm t})\frac{1}{2}B}{1+\frac{B}{2}}=E_B,
\end{eqnarray}
where $\epsilon_{\infty}$ is the prescribed high frequency permittivity of $\epsilon_{\rm P}(\omega)$ and
$\Omega=\omega_{\rm d}[1-\frac{B}{2},1+\frac{B}{2}]$ the approximation domain, where $\omega_{\rm d}$ is the center frequency 
and $B$ the relative bandwidth with $0<B<2$. 
This physical bound has been derived solely based on the assumption of linearity, time-translational invariance and passivity by employing
the theory for Herglotz functions (or equivalently the so-called Positive Real functions) and its associated sum rules \cite{Gustafsson+Sjoberg2010a}. 
It should be noted that the bound \eref{eq:sumruleconstraint} is an ultimate physical bound that must be satisfied by any passive material,
but it is not necessarily achievable (tight). To this end, the convex optimization approach \eref{eq:cvxdef} provides a complementary technique to study
the realizability of such physical bounds, \cf \cite{Nordebo+etal2014b}.


\begin{figure}[htb]
\begin{picture}(50,140)
\put(90,0){\makebox(50,130){\includegraphics[width=7.5cm]{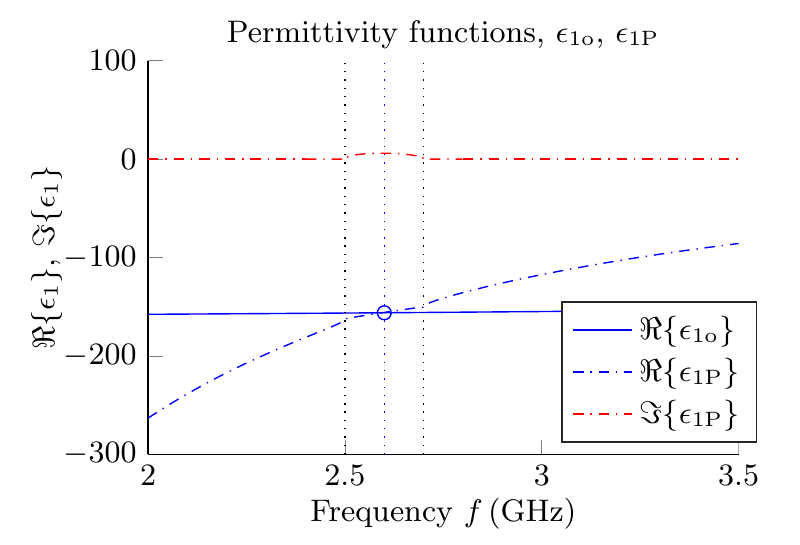}}} 
\end{picture}
\caption{Permittivity functions corresponding to the conjugate match $\epsilon_{\rm 1o}$, the tuned Drude model $\epsilon_{\rm 1D}$ and the optimized passive model $\epsilon_{\rm 1P}$.
The surrounding medium is a saline solution with conductivity $\sigma=1$\unit{S/m} and the approximation domain is $\Omega=[2.5,2.7]$\unit{GHz}.
The constant target permittivity is $\epsilon_{\rm t}=-156.1\approx\Re\!\left\{\epsilon_{\rm 1o}(\omega)\right\}$ for $\omega\in\Omega$.}
\label{fig:matfig92m}
\end{figure}


\begin{figure}[htb]
\begin{picture}(50,90)
\put(90,0){\makebox(50,80){\includegraphics[width=7.5cm]{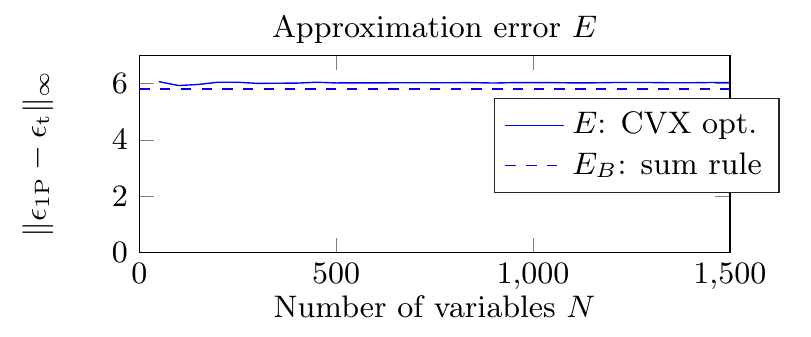}}} 
\end{picture}
\caption{Approximation error $E=\|\epsilon_{\rm 1P}-\epsilon_{\rm t}\|_\infty$ as a function of the number of approximating basis functions $N$.
The surrounding medium is a saline solution with conductivity $\sigma=1$\unit{S/m} and the approximation domain is $\Omega=[2.5,2.7]$\unit{GHz}.
The constant target permittivity is $\epsilon_{\rm t}=-156.1\approx\Re\!\left\{\epsilon_{\rm 1o}(\omega)\right\}$ for $\omega\in\Omega$.
Here,  $B=0.0769$ and $E_B=5.8185$.}
\label{fig:matfig100m}
\end{figure}

To illustrate the bandwidth limitation associated with such metamaterials, it is shown 
in \Fref{fig:matfig92m} the corresponding optimization results when the target permittivity $\epsilon_{\rm 1o}(\omega)$ has been replaced for 
the constant real negative value $\epsilon_{\rm t}=-156.1\approx\Re\!\left\{\epsilon_{\rm 1o}(\omega)\right\}$ for $\omega\in\Omega$.
In \Fref{fig:matfig100m} is shown the corresponding approximation error $E$ as a function of the number of basis functions $N$ in a comparison to
the ultimate lower bound \eref{eq:sumruleconstraint}.
It is noted that the realizable error $E$ is only slightly larger than the ultimate physical bound $E_B$ in \eref{eq:sumruleconstraint}.

The following interesting features are observed regarding the approximation of metamaterials in this particular numerical example.
\begin{itemize}
\item The problem to approximate the conjugate match $\epsilon_{\rm 1o}$ over a finite frequency band $\Omega$ as in section \ref{sect:exoptdispmodeling} is essentially the same as
to approximate a metamaterial with a constant complex value $\epsilon_{\rm t}= -156.1 + \iu 37.61 \approx\epsilon_{\rm 1o}(\omega)$ for $\omega\in\Omega$. 
This is a situation in which the result \eref{eq:sumruleconstraint} cannot be straightforwardly extended but convex optimization \eref{eq:cvxdef} can
be used to study the realizability and the bandwidth capabilities of such a hypothetical passive material.
\item Whereas the approximation of a metamaterial with a constant real negative permittivity over a finite bandwidth implies severe bandwidth limitations,
the presence of an additional constant imaginary part in the target permittivity implies that the bandwidth limitation is no longer severe, and there exist
a hypothetical passive (realizable) permittivity function that can approximate the metamaterial (the conjugate match) with a very small approximation error over the whole approximation domain $\Omega$,
as illustrated in Figures \ref{fig:matfig92} through \ref{fig:matfig100}.
\end{itemize}

\section{Summary}
A study on the physical limitations for radio frequency absorption in gold nanoparticle (GNP) suspensions has been presented in this paper.
The analysis is based on classical electromagnetic theory regarding the energy absorption in small spherical particles in a lossy medium (\cf general Mie theory)
and is extended here to take the skin effect of the surrounding medium into account. The main result is the derivation of an
optimal conjugate match with respect to the surrounding medium giving a physical limitation for the best possible absorption inside the spherical suspension.
As such, the theoretical result defines an optimum plasmonic resonance, which can furthermore be put into the framework of approximating a metamaterial
over a given bandwidth.
The heating is quantified by a relative heating coefficient, and it is demonstrated that for a surrounding medium consisting of
a weak electrolyte solution a significant RF-heating can be achieved inside the GNP suspension, provided that an electrophoretic 
mechanism described by the Drude model can be tuned into resonance at the desired frequency.

\ack
This work was supported by the Swedish Foundation for Strategic Research (SSF).

\appendix
\section{Vector spherical waves}\label{sect:spherical}
The definition of the vector spherical waves together with some important results that are not readily available in the literature
are presented in this appendix. In particular, we supply explicit formulas for the transition matrices $t_{\tau l}$ and $r_{\tau l}$
defined in \eref{eq:btaumldef} and \eref{eq:ataumldef} below, as well as the energies (as squared integrals) of the 
spherical Bessel functions \eref{eq:firstLommelcomplxarg} and the regular vector waves \eref{eq:W1ldef1} and \eref{eq:W2ldef}
in the case with lossy materials. The squared integrals are based on the first Lommel integral in the case when the wavenumber $k$ is complex valued.

\subsection{Definition of vector spherical waves}\label{sect:sphericaldef}
In a source-free region the electromagnetic field can be expanded in vector spherical waves as
\begin{eqnarray}\label{eq:Esphdef}
\bm{E}(\bm{r})=
\sum_{l,m,\tau}a_{\tau ml}{\bf v}_{\tau ml}(k\bm{r})+b_{\tau ml}{\bf u}_{\tau ml}(k\bm{r}),  \\
\bm{H}(\bm{r}) 
=\frac{1}{\iu\eta_0\eta}
\sum_{l,m,\tau}a_{\tau ml}{\bf v}_{\bar{\tau} ml}(k\bm{r})+b_{\tau ml}{\bf u}_{\bar{\tau} ml}(k\bm{r}),  \nonumber
\end{eqnarray}
where ${\bf v}_{\tau ml}(k\bm{r})$ and ${\bf u}_{\tau ml}(k\bm{r})$ are the regular and the outgoing vector spherical waves, respectively, 
and $a_{\tau ml}$ and $b_{\tau ml}$ the corresponding multipole coefficients,
see \eg \cite{Bostrom+Kristensson+Strom1991,Arfken+Weber2001,Jackson1999,Newton2002}.
Here, $l=1,\ldots,\infty$, $m=-l,\ldots,l$ and $\tau=1,2$, where
$\tau=1$ indicates a transverse electric (\textrm{TE}) magnetic multipole and $\tau=2$ a transverse magnetic (\textrm{TM}) electric multipole,
and $\bar{\tau}$ denotes the complement of $\tau$, \ie $\bar{1}=2$ and $\bar{2}=1$.

The solenoidal (source-free) regular vector spherical waves are defined here by
\begin{eqnarray}\label{eq:vdef}
\displaystyle{\bf v}_{1 ml}(k{\bm{r}})  =   \frac{1}{\sqrt{l(l+1)}}\nabla\times({\bm{r}}{\rm j}_l(kr){\rm Y}_{ml}(\hat{\bm{r}})) \\
\qquad =   {\rm j}_l(kr){\bf A}_{1 ml}(\hat{\bm{r}}), \nonumber \\
{\bf v}_{2 ml}(k\bm{r})   =   \displaystyle \frac{1}{k}\nabla\times{\bf v}_{1 ml}(k\bm{r}) \\
 \qquad =\displaystyle\frac{(kr{\rm j}_l(kr))^{\prime}}{kr}{\bf A}_{2 ml}(\hat{\bm{r}})+\sqrt{l(l+1)}\frac{{\rm j}_l(kr)}{kr}{\bf A}_{3 ml}(\hat{\bm{r}}), \nonumber
\end{eqnarray}
where ${\rm Y}_{ml}(\hat{\bm{r}})$ are the spherical harmonics, ${\bf A}_{\tau ml}(\hat{\bm{r}})$ the vector spherical harmonics and ${\rm j}_l(x)$ the spherical Bessel functions of order $l$,
\cf \cite{Bostrom+Kristensson+Strom1991,Arfken+Weber2001,Jackson1999,Newton2002,Olver+etal2010}. 
Here, $(\cdot)^\prime$ denotes a differentiation with respect to the argument of the spherical Bessel function.
The outgoing (radiating) vector spherical waves ${\bf u}_{\tau ml}(k{\bm{r}})$ are obtained by replacing
the regular spherical Bessel functions ${\rm j}_l(x)$ above for the spherical Hankel functions of the first kind, ${\rm h}_l^{(1)}(x)$, 
see \cite{Bostrom+Kristensson+Strom1991,Jackson1999,Olver+etal2010}.

The vector spherical harmonics ${\bf A}_{\upsilon lm}(\hat{\bm{r}})$ are given by
\begin{eqnarray}\label{eq:Adef}
{\bf A}_{1ml}(\hat{\bm{r}})  =   \displaystyle\frac{1}{\sqrt{l(l+1)}}\nabla\times\left( \bm{r}{\rm Y}_{ml}(\hat{\bm{r}}) \right),   \\
{\bf A}_{2ml}(\hat{\bm{r}})  =  \hat{\bm{r}}\times{\bf A}_{1ml}(\hat{\bm{r}}),  \\
{\bf A}_{3ml}(\hat{\bm{r}}) = \hat{\bm{r}}{\rm Y}_{ml}(\hat{\bm{r}}), 
\end{eqnarray}
where $\upsilon=1,2,3$, and where the spherical harmonics ${\rm Y}_{ml}(\hat{\bm{r}})$ are given by
\begin{eqnarray}
{\rm Y}_{ml}(\hat{\bm{r}})=(-1)^m\sqrt{\frac{2l+1}{4\pi}}\sqrt{\frac{(l-m)!}{(l+m)!}}{\rm P}_{l}^m(\cos\theta)\eu^{{\rm i}m\phi}, \nonumber
\end{eqnarray}
and where ${\rm P}_{l}^m(x)$ are the associated Legendre functions \cite{Arfken+Weber2001,Jackson1999,Olver+etal2010}.
The vector spherical harmonics are orthonormal on the unit sphere, and hence
\begin{equation}\label{eq:Aorthonormal}
\int_{\Omega_0}{\bf A}_{\upsilon ml}^*(\hat{\bm{r}})\cdot{\bf A}_{\upsilon^\prime m^\prime l^\prime }(\hat{\bm{r}}){\rm d}\Omega
=\delta_{\upsilon\upsilon^\prime}\delta_{mm^\prime}\delta_{ll^\prime},
\end{equation}
where $\Omega_0$ denotes the unit sphere and ${\rm d}\Omega=\sin\theta{\rm d}\theta{\rm d}\phi$.

\subsection{Transition matrices for a homogeneous sphere}\label{sect:sphericalT}

Consider the scattering of an electromagnetic field due to a homogeneous sphere of radius $r_1$, permittivity $\epsilon_1$, permeability $\mu_1$
and wavenumber $k_1=k_0\sqrt{\mu_1\epsilon_1}$.
The medium surrounding the sphere is characterized by the permittivity $\epsilon$, permeability $\mu$ and wave number $k=k_0\sqrt{\mu\epsilon}$.
The incident and the scattered fields for $r>r_1$ are expressed as in \eref{eq:Esphdef} with multipole coefficients $a_{\tau ml}$ and $b_{\tau ml}$, respectively,
and the interior field is similarly expressed using regular vector spherical waves for $r<r_1$ with multipole coefficients $a_{\tau ml}^{(1)}$.
By matching the tangential electric and magnetic fields at the boundary of radius $r_1$, it can be shown that
\begin{eqnarray}\label{eq:btaumldef}
b_{\tau ml}=t_{\tau l}a_{\tau ml}, \\
a_{\tau ml}^{(1)}=r_{\tau l}a_{\tau ml},\label{eq:ataumldef}
\end{eqnarray}
where $t_{\tau l}$ and $r_{\tau l}$ are transition matrices for scattering and absorption given by
{ \scriptsize
\begin{eqnarray} 
t_{1l}=   
\displaystyle \frac{\textrm{j}_l(kr_1) (k_1r_1\textrm{j}_l(k_1r_1))^\prime \mu -\textrm{j}_l(k_1r_1) (kr_1\textrm{j}_l(kr_1))^\prime \mu_1}
{\textrm{j}_l(k_1r_1) (kr_1\textrm{h}_l^{(1)}(kr_1))^\prime \mu_1-\textrm{h}_l^{(1)}(kr_1) (k_1r_1\textrm{j}_l(k_1r_1))^\prime \mu }, \nonumber \\
t_{2l} =  
 \displaystyle  \frac{ \textrm{j}_l(k_1r_1) (kr_1\textrm{j}_l(kr_1))^\prime \epsilon_1 - \textrm{j}_l(kr_1) (k_1r_1\textrm{j}_l(k_1r_1))^\prime \epsilon}
{ \textrm{h}_l^{(1)}(kr_1) (k_1r_1\textrm{j}_l(k_1r_1))^\prime\epsilon- \textrm{j}_l(k_1r_1) (kr_1\textrm{h}_l^{(1)}(kr_1))^\prime\epsilon_1}, \nonumber \\
r_{1l} = 
\displaystyle  \frac{\textrm{j}_l(kr_1) (kr_1\textrm{h}_l^{(1)}(kr_1))^\prime \mu_1-\textrm{h}_l^{(1)}(kr_1) (kr_1\textrm{j}_l(kr_1))^\prime \mu_1}
{\textrm{j}_l(k_1r_1) (kr_1\textrm{h}_l^{(1)}(kr_1))^\prime \mu_1-\textrm{h}_l^{(1)}(kr_1) (k_1r_1\textrm{j}_l(k_1r_1))^\prime \mu }, \nonumber \\
r_{2l} = 
 \displaystyle  -\frac{ (\textrm{j}_l(kr_1) (kr_1\textrm{h}_l^{(1)}(kr_1))^\prime-\textrm{h}_l^{(1)}(kr_1) (kr_1\textrm{j}_l(kr_1))^\prime)\sqrt{\epsilon} \sqrt{\epsilon_1} \sqrt{\mu_1}}
   { ( \textrm{h}_l^{(1)}(kr_1) (k_1r_1\textrm{j}_l(k_1r_1))^\prime\epsilon- \textrm{j}_l(k_1r_1) (kr_1\textrm{h}_l^{(1)}(kr_1))^\prime\epsilon_1)\sqrt{\mu }}.\nonumber 
\end{eqnarray}}

\subsection{First Lommel integral for spherical Bessel functions with complex-valued argument}\label{sect:Lommel}
The first Lommel integral  is given by
\begin{eqnarray}\label{eq:Lommel1}
\int {\rm C}_{\nu}(a\rho){\rm D}_\nu(b\rho)\rho{\rm d}\rho \\
\qquad =\frac{\rho\left(a{\rm C}_{\nu+1}(a\rho) {\rm D}_{\nu}(b\rho) -b{\rm C}_{\nu}(a\rho) {\rm D}_{\nu+1}(b\rho) \right)}{a^2-b^2}, \nonumber
\end{eqnarray}
where $a$ and $b$ are complex-valued constants and ${\rm C}_{\nu}(\cdot)$ and ${\rm D}_{\nu}(\cdot)$ are arbitrary cylinder functions, \ie 
the Bessel function, the Neumann function, the Hankel functions of the first and second kind
${\rm J}_{\nu}(\cdot)$, ${\rm Y}_{\nu}(\cdot)$, ${\rm H}_{\nu}^{(1)}(\cdot)$, ${\rm H}_{\nu}^{(2)}(\cdot)$, respectively, or any nontrivial linear combination
of these functions, see 10.22.4 and 10.22.5 in \cite{Olver+etal2010},  and pp.\ 133--134 in \cite{Watson1995}.

Let $a=\kappa$ and $b=\kappa^*$, where $\kappa\neq\kappa^*$, \ie $\kappa$ is not real valued, and consider the case
\begin{equation}\label{eq:CLomdef}
{\rm C}_\nu(\kappa\rho)=A {\rm J}_\nu(\kappa\rho)+B{\rm H}_\nu^{(1)}(\kappa\rho),
\end{equation}
where $A$ and $B$ are complex-valued constants. Let
\begin{equation}\label{eq:DLomdef1}
{\rm D}_{\nu}(\kappa^*\rho)={\rm C}_\nu^*(\kappa\rho)=A^* {\rm J}_\nu(\kappa^*\rho)+B^*{\rm H}_\nu^{(2)}(\kappa^*\rho),
\end{equation}
where the conjugate rules ${\rm J}_\nu^*(\zeta)={\rm J}_\nu(\zeta^*)$ and ${{\rm H}_\nu^{(1)}}^*(\zeta)={\rm H}_\nu^{(2)}(\zeta^*)$
have been used, \cf \cite{Olver+etal2010}.
The first Lommel integral \eref{eq:Lommel1} now yields
\begin{equation}\label{eq:Lommel1b}
\int |{\rm C}_\nu(\kappa\rho)|^2\rho{\rm d}\rho=
\frac{\rho\Im\!\left\{\kappa{\rm C}_{\nu+1}(\kappa\rho) {\rm C}_\nu^*(\kappa\rho) 
 \right\}}{\Im\!\left\{\kappa^2\right\}}.
\end{equation}

The spherical Bessel, Neumann and Hankel functions of the first and second kind are given by
${\rm j}_l(\zeta)=\sqrt{\frac{\pi}{2\zeta}}{\rm J}_{l+1/2}(\zeta)$, ${\rm y}_l(\zeta)=\sqrt{\frac{\pi}{2\zeta}}{\rm Y}_{l+1/2}(\zeta)$,
${\rm h}_l^{(1)}(\zeta)=\sqrt{\frac{\pi}{2\zeta}}{\rm H}_{l+1/2}^{(1)}(\zeta)$ and ${\rm h}_l^{(2)}(\zeta)=\sqrt{\frac{\pi}{2\zeta}}{\rm H}_{l+1/2}^{(2)}(\zeta)$, 
respectively, \cf \cite{Olver+etal2010}. An arbitrary linear combination of spherical Bessel and Hankel functions can hence be written as
\begin{equation}
{\rm s}_l(kr)=A {\rm j}_l(kr)+B{\rm h}_l^{(1)}(kr)=\sqrt{\frac{\pi}{2kr}}{\rm C}_{l+1/2}(kr),
\end{equation}
where ${\rm C}_{l+1/2}(kr)$ is the corresponding cylinder function as defined in \eref{eq:CLomdef}. 
The first Lommel integral for spherical Bessel functions  with complex-valued arguments can now be derived as
\begin{eqnarray}\label{eq:firstLommelcomplxarg}
\int\left|{\rm s}_l(kr)\right|^2r^2{\rm d} r
=\frac{\pi}{2|k|}\int\left|{\rm C}_{l+1/2}(kr) \right|^2r{\rm d} r \\
\quad =\frac{\pi}{2|k|}\frac{r\Im\!\left\{k{\rm C}_{l+1+1/2}(kr) {\rm C}_{l+1/2}^*(kr) \right\}}{\Im\!\left\{k^2\right\}} \nonumber \\
\quad=\frac{r^2 \Im\!\left\{k\sqrt{\frac{\pi}{2kr}}{\rm C}_{l+1+1/2}(kr) 
\left(\sqrt{\frac{\pi}{2kr}}{\rm C}_{l+1/2}(kr)\right)^* \right\}}{\Im\!\left\{k^2\right\}} \nonumber \\
\qquad=\frac{r^2 \Im\!\left\{k{\rm s}_{l+1}(kr) {\rm s}_{l}^*(kr) \right\}}{\Im\!\left\{k^2\right\}}.\nonumber
\end{eqnarray}

\subsection{Orthogonality of the regular spherical waves}\label{sect:sphericalorth}
Due to the orthonormality of the vector spherical harmonics \eref{eq:Aorthonormal},
the regular spherical waves are orthogonal over the unit sphere with
\begin{eqnarray}\label{eq:vorthogonal1}
\displaystyle\int_{\Omega_0}{\bf v}_{\tau ml}^*(k{\bm{r}})\cdot{\bf v}_{\tau^\prime m^\prime l^\prime}(k{\bm{r}}){\rm d} \Omega \\
\qquad =\displaystyle\delta_{\tau\tau^\prime}\delta_{mm^\prime}\delta_{ll^\prime}S_{\tau l}(k,r), \nonumber
\end{eqnarray}
where
\begin{eqnarray}\label{eq:Stauldef}
S_{\tau l}(k,r)=\displaystyle\int_{\Omega_0}|{\bf v}_{\tau ml}(k\bm{r})|^2{\rm d}\Omega  \\
 =\left\{\begin{array}{ll}
\displaystyle  \left|{\rm j}_{l}(kr)\right|^2 & \textrm{for}\  \tau=1,  \vspace{0.2cm} \\
\displaystyle  \left|\frac{{\rm j}_{l}(kr)}{kr}+{\rm j}_{l}^\prime(kr)\right|^2+l(l+1) \left|\frac{{\rm j}_{l}(kr)}{kr}\right|^2  & \textrm{for}\  \tau=2. \nonumber
\end{array}\right.
\end{eqnarray}
As a consequence, the regular spherical waves are also orthogonal over a spherical volume $V_{r_1}$ with radius $r_1$ yielding
\begin{eqnarray}\label{eq:vorthogonal2}
\displaystyle\int_{V_{r_1}}{\bf v}_{\tau ml}^*(k{\bm{r}})\cdot{\bf v}_{\tau^\prime m^\prime l^\prime}(k{\bm{r}}){\rm d} v \\
\qquad =\displaystyle\delta_{\tau\tau^\prime}\delta_{mm^\prime}\delta_{ll^\prime}W_{\tau l}(k,r_1), \nonumber
\end{eqnarray}
where
\begin{eqnarray}\label{eq:Wtauldef}
W_{\tau l}(k,r_1)=\int_{V_{r_1}}\left|{\bf v}_{\tau ml}(k\bm{r})\right|^2{\rm d} v \\
\qquad =\int_{0}^{r_1}S_{\tau l}(k,r)r^2{\rm d} r, \nonumber
\end{eqnarray}
where ${\rm d} v=r^2{\rm d}\Omega{\rm d} r$ and $\tau=1,2$.

For complex-valued arguments $k$, $W_{1l}(k,r_1)$ is obtained from \eref{eq:firstLommelcomplxarg} as
\begin{eqnarray}\label{eq:W1ldef1}
W_{1l}(k,r_1)=\int_{0}^{r_1}\left|{\rm j}_{l}(kr)\right|^2 r^2{\rm d} r \\
\qquad =\frac{r_1^2 \Im\!\left\{k{\rm j}_{l+1}(kr_1) {\rm j}_{l}^*(kr_1) \right\}}{\Im\!\left\{k^2\right\}}.\nonumber
\end{eqnarray}
By using the following recursive relationships
\begin{eqnarray}
\displaystyle \frac{{\rm j}_l(kr)}{kr}=\frac{1}{2l+1}\left({\rm j}_{l-1}(kr)+{\rm j}_{l+1}(kr) \right), \nonumber \\
\displaystyle {\rm j}_l^\prime(kr)=\frac{1}{2l+1}\left(l{\rm j}_{l-1}(kr)-(l+1){\rm j}_{l+1}(kr) \right), \nonumber
\end{eqnarray}
where $l=1,2,\ldots$, \cf \cite{Olver+etal2010}, it can be shown that
\begin{eqnarray}\label{eq:W2ldef}
W_{2 l}(k,r_1)\\
=\int_{0}^{r_1}\left(\left|\frac{{\rm j}_{l}(kr)}{kr}+{\rm j}_{l}^\prime(kr)\right|^2 
 +l(l+1)  \left|\frac{{\rm j}_{l}(kr)}{kr}\right|^2\right)r^2{\rm d} r \nonumber \\
\quad =\frac{1}{2l+1}\left((l+1)W_{1,l-1}(k,r_1)+lW_{1,l+1}(k,r_1) \right). \nonumber
\end{eqnarray}

\section*{References}


\end{document}